# Analysis and constructive criticism of the official data protection impact assessment of the German Corona-Warn-App

Rainer Rehak[1][0000-0002-8244-8532], Christian R. Kühne[2] and Kirsten Bock[3]

[1] Weizenbaum Institute for the Networked Society, Hardenbergstraße 32
10623 Berlin, Germany
`rainer.rehak@wzb.eu`

[2] Forum Computer Scientists für Peace and Societal Responsibility (FIfF), Goetheplatz 4
28203 Bremen, Germany

[3] Independent Centre for Privacy Protection Schleswig-Holstein (ICPP), Postbox 71 16, 24171 Kiel, Germany

On June 15, 2020, the official data protection impact assessment (DPIA) for the German Corona-Warn-App (CWA) was made publicly available. Shortly thereafter, the app was made available for download in the app stores. This analysis concludes that the initially submitted DPIA had very significant weaknesses and provides examples for this claim. However since then, the quality of the official DPIA increased immensely. To illustrate the discussion and learning curve, the initial weaknesses are documented here. First in this paper, the purpose of a DPIA is recalled in a nutshell. According to Article 35 of the GDPR, this consists primarily of identifying the risks to the fundamental rights and freedoms of natural persons. This paper documents specific methodological, technical and legal shortcomings of the initial DPIA of the CWA: 1) Its focus only lies on the app itself, neither on the whole processing procedure nor on the infrastructure used. 2) It only briefly touches on the main data protection specific attacker, the processing organization itself. And 3) The discussion of effective safeguards to all risks such as the ones posed by Google and Apple are insufficiently worked out. This paper then documents the constructive comments and suggestions uttered before and after the initial release. As of now, some of those constructive contributions have been worked into the current DPIA, such as 1) and 2), but some central ones haven't, such as 3).

**Keywords:** Data protection, data protection impact assessment, DPIA, corona apps, CWA, digital contact tracing, decentralisation, GDPR, privacy

## 1 Introduction

After Germany opted for a data protection friendly, decentralized approach to automated, app-based digital contact tracing, the Corona-Warn-App (CWA) [1] was released by the Robert Koch-Institute on June 16, 2020. A technically outstanding system was created within a short period of time. It uses current software frameworks, is open source and was partly created with transparent participatory work processes building up a community. In the process, publicly formulated criticism as well as contributions from the interested (specialist) public were taken into account in many cases. Now it remains to be hoped that a new standard has been established for future governmental IT projects in Germany. However, the CWA is not without data protection issues [16]. The CWA's publication was accompanied by a comprehensive Data Protection Impact Assessment (DPIA) [15], which is mandatory according to Art. 35 GDPR and which recognizes many critical data protection issues of the CWA.

### 1.1 Function and architecture of the CWA

Main purpose of the CWA is to warn individuals who have had contact with infected persons in the past so they can voluntarily self-quarantine. Later on the functions of storing vaccination and testing certificates were added, but this is omitted here. The warning functionality is put to practice by the smartphone sending regularly changing strings (pseudonymous temporary identifiers, tempIDs) via Bluetooth at regular intervals using the "Bluetooth Low Energy Beacons" (BTLE) standard, and at the same time receiving the temporary identifiers (tempIDs) from other apps accordingly, when they are in close vicinity [16]. Hence, each app keeps two buckets with the tempIDs of the last fourteen days, one bucket for the tempIDs sent and one for the tempIDs received. Actual location information, for example GPS data, is not processed or even collected by this system. The underlying Bluetooth functionality is managed by the Exposure Notification Framework (ENF or GAEN) provided by Apple and Google in their respective mobile operating systems [13].

In case of positive testing, only the temporary identifiers – the daily seeds, to be precise – sent out by the person during the past 14 days are uploaded to the CWA server. Those uploaded temporary identifiers indicate infectiousness. If any other app has received those tempIDs, it means that there was a possibly relevant contact event. Therefore the other apps regularly download the current data set of all infection-indicating tempIDs and check for matches, i.e. if they have seen any of those tempIDs. If yes, they calculate locally on the smartphone whether there is a risk of infection based on the duration and proximity of the contact, as well as the state of illness of the infected person at the time of contact. If there is a risk, the user is then warned accordingly by the app (not by the server). Since the server only knows the ever changing tempIDs of infected users, it can neither create a contact history nor calculate the social network of all users [16]. Therefore, this decentralized variant is much more data protection-friendly, yet also more traffic-intensive than a centralised one, e.g. TousAntiCOVID in France or TraceTogether in Signapore.

### 1.2 Data protection impact assessment (DPIA)

In order to find, analyze and discuss the diverse consequences of this European project for large-scale contact tracing under government responsibility the European General Data Protection Regulation (GDPR) provides the instrument of a data protec-tion impact assessment, which in specific cases is even mandatory [16]. Concretely Art. 35 GDPR ("Data Protection Impact Assessment", DPIA) states [3]:

*(1) Where a type of processing in particular using new technologies, and taking into account the nature, scope, context and purposes of the processing, is likely to result in a high risk to the rights and freedoms of natural persons, the controller shall, prior to the processing, carry out an assessment of the impact of the envisaged processing operations on the protection of personal data. A single assessment may address a set of similar processing operations that present similar high risks.*

*[…]*

*(7) The assessment shall contain at least*

*(a) a systematic description of the envisaged processing operations and the purposes of the processing, including, where applicable, the legitimate interest pursued by the controller;*

*(b) an assessment of the necessity and proportionality of the processing operations in relation to the purposes;*

*(c) an assessment of the risks to the rights and freedoms of data subjects referred to in paragraph 1; and*

*(d) the measures envisaged to address the risks, including safeguards, security measures and mechanisms to ensure the protection of personal data and to demonstrate compliance with this Regulation taking into account the rights and legitimate interests of data subjects and other persons concerned.*

It is essential for a DPIA according to the GDPR that the focus does not lie on the technology itself, in this case the Corona-Warn-App, but instead the DPIA should focus on the processing as a whole, which consists of several series of processing activities which can be, in part, supported by technology like an app [16]. All considerations must therefore go beyond the use of "the app" and embrace the whole process including servers, network infrastructure or operating system frameworks and even the process parts without technology use. So to summarize it, the boundary of the app is not the boundary of the processing [16].

### 1.3    Official handling of conducting the DPIA

As described in detail below, the initial DPIA had several weaknesses but the official DPIA working group later on explicitly responded to several critical contributions to the discussion on centralized or decentralized approaches, as well as to the model DPIA presented by the Forum Computer Scientists for Peace and Societal Responsibility on April 14, 2020 [2]. By making the official CWA DPIA publicly available, the societal significance of the CWA system was acknowledged: The CWA might accustom individuals to be continuously monitored directly in their day-to-day life by a governmental IT system. One way to take up and deal with concerns like that is precisely through a public debate based on a high quality data protection impact assessment. In this respect, DPIAs contribute to the systematic public discussion of the surveillance and control aspects of the digitized society. The initial CWA DPIA has correctly identified some risk. For example, the problematic role of the Exposure Notifi-cation Framework (ENF) of Google and Apple was recognized and analyzed up to the realization "that they jointly developed the ENF according to their ideas and integrated it as a separate system function in their respective operating systems; the storage period of tag keys […], the configuration parameters […] and the availability of the ENF are unilaterally determined by Google and Apple. Apps may only access the functions and data of the ENF if unilateral specifications by Apple or Google are met. To that extent, Apple and Google determine the purpose and essential means of processing by the ENF." (Section 8.8.3 in the initial DPIA [15]) The identification feature "IP address" was also given the essential relevance it deserves, which is often misappropriated elsewhere. From a data protection point of view, it is generally not a matter of obtaining the plain name of a person on the basis of IP addresses in order to then identify that person. The IP address itself is already the identification and therefore personal data, as has also been established in the relevant rulings of the ECJ of 2016 and the German Federal Court of Justice in 2017. In this respect, it is correctly stated in the official DPIA "Insofar and as long as the RKI stores or otherwise processes anonymous data in connection with an IP address on its own, it is therefore personal data for the RKI as a whole." (Section 10.1.1 [15]). The CWA-DPIA even goes beyond the previous analyses and findings of mid-April 2020 [2] with regard to the nature of the data at various points of the processing, when it states, for example: "The list of positive keys of other users downloaded from the [CWA server], which are processed locally on the user's smartphone, are health data for the RKI as long as these data are on the [CWA server], since they indicate a coronavirus infection of the persons behind the respective positive key or the (former) day keys" (Section 10.1.3 [15]). All these points have been aptly elaborated, but criticism must nevertheless also be levelled. In the following, however, we will not undertake a detailed analysis of the initial CWA-DPIA, but will highlight only a few, but essential and critical aspects of the CWA-DPIA with regard to methodological, technical and legal deficits. First, we will list general points of criticism, after which we will address specific points and finally propose improvements.

## 2    General points of criticism

The initial DPIA report contained many shortcomings, although DPIAs should be part of the standard repertoire of every personal data processing project since May 2016 due to the new obligations of the GDPR. The risks identified already in mid-April by an independent expert group in their own DPIA [2] are still valid for all decentralised implementations. The official CWA had many methodological weaknesses and did not follow a systematic approach to fulfill the requirements of the GDPR. Instead of systematically transforming the normative requirements of data protection law into functional requirements (see the standard data protection model [7]), the initial official DPIA apparently drew the problem definitions from general knowledge on IT security. Orientation towards the relevant guidelines of the European Data Protection Board, in particular on consent (5/2020), was missing. There are three substantial deficiencies which entail many further flaws in the risk analysis of a DPIA. First, the GDPR - and thus a DPIA - never refers to only one selected technical component, but always to the processing activity as a whole. Secondly, in the official DPIA there was no consistent data protection-specific attacker model, which systematically focuses on the fundamental rights infringements by the controllers (and the technical operators commissioned by them). And thirdly, the legal constellations and responsibilities regarding risk minimisation were not sufficiently presented and assessed.

## 2.1 Protective function of a DPIA

The function of a DPIA according to Art. 35 GDPR is to make the risks of a data processing for fundamental rights of natural persons primarily visible for the controller - we generically adopt "the controller" as a general role designation from the GDPR - itself. This is often provocative because the controller is analytically considered the main attacker for the rights and freedoms of data subjects [16]. The construction is daring: the controller, who is considered the main aggressor, is at the same time the one who is supposed to determine and then implement measures with which these risks for the data subjects can be reduced to a responsible level. However, this is at the same time the reason for the standardisation of protective provisions by data protection law and for the requirement of a DPIA for particularly risky processing activities. A DPIA report should enable the responsible party to ensure that measures are taken to effectively reduce the identified risks to a responsible level. However, it must also explicitly show when significant risks cannot be reduced. The latter may have the consequence that a processing activity, measured against the requirements of the GDPR, cannot be implemented as planned and is therefore impermissible. A dismissive discussion of risks - in the present DPIA report particularly clear in the context of the handling of the operating system functions provided by Google and Apple, and the reference to the fact that the controller is not planning any protective measures here - then completely misses the point of a DPIA. On the contrary, a DPIA must call for helpful recommendations for protective measures with regard to identified risks, or it must show that the risks remain untreated, i.e. open problems and untreated risks must be named as such. With regard to the problem of the existing blatant dependencies on the manufacturers of smartphone operating systems, it may well be concluded that these proprietary functions should not be used due to the lack of sufficient verifiability of the associated processing on the part of the manufacturers. Please note that the first paragraph of a section or subsection is not indented. The first paragraphs that follows a table, figure, equation etc. does not have an indent, either.

## 2.2 Requirements and methodology of a DPIA

Article 35 of the GDPR specifies the requirements for a DPIA. The basis is a description of the processing activity, whereby the description necessarily accesses a documentation of the properties of all components used. The methods and guidelines that can be used for this purpose are listed in the FIfF-DPIA starting on p. 12. The main methodological flaw of the official CWA DPIA is that only the functions of the app are considered, but not the processes of the entire processing with all its data, all IT components used and processes (partially illustrated in the overview of the CWA architecture, Fig. 2, section 8.1 [15]). The integration of the verification hotline into the DPIA points in the right direction, but this could also have been done more consistently (see e.g. Section 10.2.3.5 [15]). The scope of a DPIA is processing (according to Art. 4(2) GDPR), not just an IT component. In this CWA report, although essential components, such as the server operation are named in Figures 2 and 16, their functionalities are only roughly outlined. The data flows and data stocks up to the health authorities and the doctors involved would have to be presented in detail, including the legal relationships of all parties involved with each other and ultimately always with reference to the responsible person. A data protection risk analysis with integrity would, for example, have placed great emphasis on the representation of the server(s) to which the tag keys of Corona infected persons are uploaded. It is precisely this high-risk point that the FIfF-DPIA has insistently drawn attention to, because the entire data protection risk depends on the degree of personal reference of the infection-indicating data. At any rate, it currently remains non-transparent which data protection-relevant characteristics the servers involved have, which transactions and data are logged, evaluated and deleted there. A vague statement that "on the part of the RKI, it is planned to delete the IP address from the server log files on the CWA server and CDN-Magenta immediately after responding to a request" and therefore "the personal reference described above in connection with an IP address would only exist for the RKI for a 'technical second'" (Section 10.1.1 [15]) is far from sufficient, because this point or this moment in the entire processing chain is the most sensitive point, since the tag keys of infected persons are personalised here via IP address. A promise of the responsible persons is not a protection measure, in front of the responsible persons. More detailed explanations follow later in the document. Moreover, a DPIA as such should not be considered or created as a "living document" (Section 1 [15]). A DPIA report claims a certain closure of the analysis and findings and concludes with concrete recommendations for risk mitigation. Nevertheless, the controller of the processing activity must of course be able to react to further changes in the context of the processing activity. This, however, is the function of a data protection management. This means that a DPIA has to be handed over to a data protection management, in this case to the hopefully actually existing data protection management system, and this means more and different than just the appointment of a data protection officer, at the RKI.

### 2.3  The processing activity

For the description of the processing activity as a systemic context, it is recommended to a) describe the sequence of purpose, purpose description, purpose separation and purpose limitation, as proposed by the SDM V2, and b) to orientate oneself at least on the 14 subprocesses that are listed as components of a processing in Art. 4 No. 2 of the GDPR. The presentation of the functional properties essential under data protection law can then be made along the risks, formed from the principles of Article 5 GDPR, or in the case of a primarily functional orientation, from the compact assurance objectives of the SDM. In concrete terms, orienting oneself not to the procedure, but mainly to the CWA app, has then led to the fatal assumption in Section 10.1 [15], namely that the data processed locally or "offline" on the smartphone are not considered part of the controller's processing activity. The CWA app is part of the process insofar as it contributes to the achievement of the purpose. It is also a product of technical design by or on behalf of the controller and thus determines in principle the possible consequences of the use of technology. Last but not least, the responsible party controls and manages the patch and update management process of the CWA app (via the provision of new signed software versions in the "App or Play Store"), in which the CWA users participate, and thus also influences the procedure and its consequences in the future. This would have been evident in a procedure-oriented analysis as opposed to an application-oriented analysis. The consequences to be considered downstream here would then not only have to deal with the risk of de-anonymisation by the controller itself (Section 10.1.2 [15]), but also with risks due to software errors or operating system functions used, which could, for example, lead to de-anonymisation by third parties. To ensure that this does not happen, a level of protection appropriate to the risk must also be guaranteed on the smartphones (Art. 32(1) GDPR). Although the DPIA understands "Corona-WarnApp" to mean not only the app itself, but also the CWA servers in many places, there is no mention of the servers. However, this area in particular is sensitive and relevant for the (non-existent) attacker model. Because apart from attacks by third parties, it is the server operatorsï themselves who can relatively easily influence the processing on the server. For example, it would have been necessary to explain how the CWA server differs from the other servers (test result server, portal server, verification server, CDN) with respect to the risks posed by their function. Furthermore, there is no systematic description of interfaces or communication relationships as well as their purpose, type of information transferred, types of access and associated protection measures. This must then show, above all, the extent to which data is forwarded to other (joint) controllers or for commissioned data processing, e.g. from or to doctors' practices, laboratories, SAP/telecom data centres or operating system manufacturers, in order to identify risky points for pontential misappropriation (in accordance with the protection goal of non-linkability and confidentiality). They are of "crucial importance for the legal accountability, controllability and auditability of data flows" (SDM V2, p. 39).

## 2.4    Risk modeling

Another conceptual methodological flaw is the lack of data protection-specific risk modeling. This is obviously due to the general lack of sufficient orientation in operational data protection. From a data protection perspective, the controller itself is considered the main aggressor on the rights and freedoms of natural persons; the principles from Article 5 GDPR then form the criteria with which risks are to be observed and assessed. This methodical approach has been considered good DPIA practice ("state of the art") since 2017 at the latest among those who know how to distinguish operational data protection from IT security issues. You can and must know this by now if you are to set up such a project. The specific data protection risk for CWA users is that the encroachment on fundamental rights by the controller and "his" data processing is too intensive and the data protection principles are not fulfilled. In concrete terms, this means that, for example, the confidentiality, integrity and purpose limitation of data processing are not adequately secured at any point in the entire processing chain - and not just on the smartphone itself - and that there are no audit and test options (transparency) for identifying whether the protective measures are actually effective and demonstrably function securely. And this is for the protection of those affected, which does not correspond to the view of IT security. Therefore, such necessary assurances as in the glossary (section 4.1 [15]) are too vague and useless when it states, for example, that "the encounter record can also be actively deleted as a whole by the CWA user at any time. The data in the operating system (collected and own day keys) are not deleted, but remain stored for 2 weeks in the protected operating system memory." At this point, the ability to control by the CWA users is removed and another transparency problem is added. It remains unclear what happens to the encounter records in the following. For example, can this information be uploaded to the server despite being deleted? What role does it play that the operating system storage is a specially protected area? And what does that mean exactly? Does the fact that it is a specially protected area have any impact at all on the risks to fundamental rights or on their minimisation if this area is not controllable by any of the concerned entities - RKI, CWA users, DS supervisory authorities? The performance of a DPA is, within the GDPR, again an essential requirement to implement Art. 25 GDPR. Art. 25 requires that data protection requirements are already to be taken into account in the planning phase, i.e. in particular the insights, assessments and ultimately the recommendations from the DPIA. A GDPR-compliant DPIA report can therefore never be available two days, or as in this case, ten hours before the delivery of an application and the start of the actual processing for a specific purpose. Such a report cannot then serve prudent planning but only the formal legalisation of the procedure. This time delay is also critical with regard to the inclusion of the point of view of the data subjects (Art. 35(9)), since the nature, scope and circumstances of this processing suggest a social debate in advance. Only an early publication creates publicity and thus the conditions for the inclusion of different points of view for exactly this processing activity. Passively waiting for specialist publications and media reports does not do justice to this responsibility, and Internet research in the thematic environment is unfortunately too little or does not lead to the desired results, as the critical discourse must take place on the subject matter itself. From the project management's point of view, postponing the go-live would therefore have been the right approach.

## 3      Concrete constructive criticism

### 3.1    Separation of the personal reference when uploading the positive keys

A very sensitive, if not the central, point for affected individuals exists at the moment when the tempIDs of those who tested positive are uploaded to the CWA server and thereby become positive keys. Through the metadata of the connection, specifically the IP address at the time of upload, the infected person is directly identifiable. Trust-ing in simple deletion of the corresponding entries in the log files by the operator is not sufficient in the case of such a guaranteed high risk for the affected persons (Section 10.1.1 [15]), looking back at the long history of anti-terror and security laws restricting fundamental rights. Rather, comprehensive, not only technical requirements must be imposed in order to make de-pseudonymisation or other identification of app users sufficiently difficult. This must be specifically prevented by legal, organisational and technical measures, as is explained in the recommendations in the FIfF-DPIA (see DPIA Chapter 9 - Recommendations). In organisational terms, the controller must establish a mixed structure strategically and the operators operationally in order to achieve this goal. The responsible party - i.e., the RKI - can, for example, strategically select several different operators: one operates the input nodes in the network and the other the servers on which data is stored. Operationally, the separation of personal data within organisations should be ensured by an appropriate departmental structure, separation of functions and separation of roles, etc., which enforce the informational separation of powers - i.e. functional differentiation - within the organisation. By far the most effective protection consists of a technical bundle of measures to ensure sufficient sender anonymity at the interface to the infrastructure of the operators. Here, solutions such as "distributed trusted servers" or anonymization infrastructures are of independent importance. organizations are conceivable (without anticipating the solution architecture at this point). After all, in a "trusted infrastructure", only those organizations that have no interest of their own in the data can be considered as operators. This would also provide effective protection against the obligation to hand over data, even to security authorities who simply give it a try. The omission of such a critical component of the data protection architecture could be judged as a serious ground for stopping the processing. One should not rely on fictions such as "user confidence that the operator will behave in a legally compliant manner and will only release data to law enforcement authorities if the legal requirement is met" (Design Decision D-11-1) when it comes to identifying risks. For legal protection, the separation of the personal reference can be stipulated by an accompanying law.

## 3.2 Dealing with risks at the ENF

The recourse to techniques and services of the operating systems Google Android and Apple iOS is another very central point at which it becomes architecturally particularly tricky. The associated risks are only indirectly addressed by rejecting responsibility for them. The retreat to not knowing at the very beginning of the DPIA report (ENF, p. 2) is inappropriate and does not meet the requirements of relentlessly explaining risks in detail and the responsibility that exists despite everything. The GDPR requires exactly that, taking responsibility for the functioning and data protection compliance of the entire system. Especially when it is stated on p. 43 that "the CWA App and the ENF [...] are central components of the overall CWA system". At this point, in this respect, not only methodological weaknesses, but also legal and political ones. These risks must be recognized, dealt with and at least put under legal conditions by the person in charge. Unlike in the context of IT security, these risks must not simply be accepted. At this point, a DPIA with integrity would have to ex-amine the concept of "joint responsibility" and recommend ways of structuring the CWA in a legally compliant manner. It is undoubtedly difficult to impose the legal and practical consequences on Google and Apple on an equal footing, especially if one is dependent on their technologies. But to completely forego the legal discussion and an exploration of the possibilities from the outset, as was the case here, is in any case not a solution. Of course, the results of an analysis can be very unpleasant, but the consequences for those affected are all the more so. So if "own findings about the inner workings [...] cannot be obtained because this framework is implemented for security reasons in a way that precludes an investigation", then it is not possible to simply "rely on the correctness of the processing in the frameworks and the descriptions" (Section 1 [15]). That exactly is not possible in a DPIA! Of course, a clinch situation that exists worldwide is implied here, but it must be made clear as a serious risk, as well as the fact that this risk cannot currently be actually countered in any way in an overall correct way. However, an inspection of the source code of the ENF would be the least that would even come close to a protective measure. Another avoidance approach to dealing with the risks is the protective claim that "by using an Android or iOS smartphone, users [have] expressed that they fundamentally trust these manufacturers or, in any case, have come to terms with the privacy risks associated with using a smartphone from these manufacturers for personal purposes or have otherwise adapted their usage behavior accordingly" (Section 11.2.4 [15]). However, unlike universal and basic multi-purpose operating system functions such as Wi-Fi, cellular, camera or data storage, the ENF is a highly specialized service created solely for and functioning only with official contact tracing apps, and thus necessary for CWA operation. It is therefore not a normal "infrastructure component" of the smartphone operating system, but an integral part - i.e. means - of the app and its purpose.

### 3.3 Consent and responsibility

Pursuant to Article 35(1) of the GDPR, the DPA must be carried out by the data controller. The data protection responsibility for a processing activity cannot simply be assigned to an authority or claimed "The RKI [is] the controller within the meaning of Article 4 No. 7 of the GDPR for the processing of users' personal data associated with the operation of the CWA" (Sections 6.1 and 8.8.1 [15]). A data protection responsibility is to be determined in accordance with the regulation in Art. 4 No. 7 GDPR. According to this, whoever determines the purposes and means of the processing is responsible. Something else can only apply if the purposes and means of this processing are specified by a law. Only in this case can the controller be determined or can the criteria specified in a law provide for the designation of the controller in accordance with legal requirements. Such a legal determination of responsibility has not been made so far. For all processing operations of the CWA, it would therefore have been necessary to determine specifically which body or bodies determine the purposes of the processing and who determines the means for this purpose. According to the project outline (sections 5 and 8.8.4.1 [15]), the Federal Ministry of Health has determined the purposes of the processing (warning of infection and notification of infection) and the means, namely the use of an app. In the DPIA, it would have been nec-essary to discuss which tasks with regard to the determination of the means and purposes actually pass to the RKI in the context of the operation. In the same way, it would have been necessary to determine whether a joint responsibility with Apple and Google arises through the integration of the ENF functionality and what consequences result from this (Section 8.8.3 [15]). For according to the CWA-DPIA, neither the Federal Ministry nor the RKI nor their processors have any influence on this part of the app functionality. However, it cannot be denied that a joint purpose-driven determination of the ENF has taken place, as the Federal Ministry has publicly opted for the technology offered by Apple and Google [2]. The responsibility does not require that the technical details are also determined in high resolution by the controller. However, insofar as joint responsibility is assumed, reference should have been made to Art. 26 GDPR and the resulting risks. The discussion of the responsibility of the app users and thus a reversal of the role as data subjects testifies to a weakness in understanding of data protection law. At most, it could be considered with regard to the storage and comparison of the positive keys of other users. In view of the fact that the CWA-DPIA assumes the legal basis of consent, the essential subject matter of which is precisely this processing, these statements are just as disconcerting as those that address the rejection of responsibility for the data subject data. This is because the exercise of data subjects' rights (e.g. revocation) vis-à-vis the controller cannot lead to the data subjects being held responsible (Section 8.8.3 [15]). In demarcation to the responsibility, the processors would rather have had to be determined and the risks arising from commissioned processing would have had to be discussed. In Section 10 [15], a structure more oriented towards the legal requirements would have been helpful. The basic prerequisite for an evaluation under data protection law is an understanding of Article 1 (2) of the GDPR with regard to the rights and freedoms of natural persons, the basic principles of data protection law and the terminology. When determining and dealing with the reference to persons as such, the question of the scope (as posed in Section 10.1 [15]) of the personal data is irrelevant. This only plays a role in assessing the risk posed by processing large amounts of personal data. The non-resolution of the personal link between content and transport data in the server log files (Section 10.1.1 [15]) is a risk that should also have been identified as such and at least placed under legal conditions. This is one of the main risks of possible attack scenarios (see above) and would have to be the subject of regular audits from now on. The differentiation of the data categories according to their processing location (personal data "at the RKI" in section 10.1.1 [15] and local data processing on the smartphone in section 10.1.2 [15]) makes sense in principle, but the DPIA is not the appropriate place to discuss questions of responsibility based on the data categories. Moreover, local processing (Section 10.1.2 [15]) on the apps is not outside the de facto sphere of influence of the controller, who determines the design of the technology and thus the ways in which users can, for example, express their consents through explicit actions. The users also have no influence on the technical design of the CWA. Why local data processing should involve categories of personal data at all remains unclear and should have been addressed in the description of the processing. It can be assumed that this section was actually intended to discuss the categories of personal data processed on the smartphones, apparently with the aim of subsuming the controller out of the procedure. The reference to the BfDI's concerns about denying responsibility for processing on smartphones points in this direction. These discussions fail to recognise that the reference to a person does not depend on whether this can be established by the controller at all or at any time. Rather, the requirements of data protection by design (Article 25 GDPR) require that the data protection principles be implemented effectively. Measures do not change the responsibility for the CWA procedure. Contrary to what is stated in the CWA-DPIA, according to the Breyer ruling of the ECJ (C-582/14), it is important that the person responsible "has legal means which allow him to identify the person concerned on the basis of the additional information available to that person's Internet access provider" (para. 49). The RKI as operator also has such legal means to access IP addresses of users. It is thus in principle able to trace the identifiers of the daily or positive keys back to the users. The fact that the data controller cannot attribute the data at Google and Apple does not change their personal reference, but once again underlines the joint responsibility for their processing in the procedure. Art. 5 para. 2 in conjunction with. (1)(a) of the GDPR require the controller to prove the lawfulness of the processing. Section 10.2 of the CWA DPIA [15] deals with this question. It should be noted here that the fulfilment of the requirements of a legal basis is a necessary condition for the lawfulness of the processing. However, the legal basis alone does not make processing lawful. It must be supplemented by the fulfilment of further requirements that the GDPR places on data controllers. With regard to consent as a legal basis, it should be noted that its effectiveness cannot be assumed as a general rule, but that the existence of the requirements must be proven in each individual case, i.e. for each consenting person and for each purpose. The requirements for consent are derived from Art. 6 (1) sentence 1 lit. a, 7 and 4 No. 11 GDPR. The EDSA has issued an opinion on the interpretation and application (Guidelines 05/2020 on consent under Regulation 2016/679). At this point, it should be noted that pursuant to Article 70 (1) of the GDPR, it is the task of the EDSA to provide guidance on the interpretation of the GDPR; reading and observing this guidance in the context of the DPIA on the CWA is strongly recommended. Art. 4 No. 11 requires that consent is given 1) voluntarily, 2) specifically, 3) informed and 4) with an unambiguous expression of will in the form of a statement or unambiguous affirmative act. A distinction must be made between consent as a legal basis with regard to the procedure and consent to the processing of special categories of data pursuant to Art. 9 (2) lit a GDPR. The DPIA also lacks the necessary level of detail of the presentation and risk discussion here. Particularly in the case of a processing of health data using a tracing technology that is new for these purposes, a detailed discussion would have been necessary. A DPIA is also not just about repeating legally prescribed criteria, but requires a discussion of the requirements with regard to ensuring their fulfilment, i.e. with an operationalisation of normative requirements into functional requirements. For example, it would have been necessary to elaborate on what information about the procedure and its purposes is required for informed consent, what risks may arise, and references to how this has been implemented in the CWA or how the risks are addressed. A reference to the fact that "it is not evident that this information could not be reliably conveyed to the user in advance of granting consent" (Section 10.2.3.2 [15]) is by no means sufficient for this. It is not possible to conclude from an action of the user that he or she is informed. Reference should also be made to the EDSA document regarding voluntariness. In particular, the CWADPIA lacks a discussion of the fact that use of the app when receiving a warning is associated with a symptomindependent testing option; this option is not afforded to non-users of the CWA. The indication that the legislator is not currently planning to make use of the app compulsory or to make it a prerequisite for relaxation is only partially suitable for a risk assessment, also because members of the Bundestag and the Landtag have already publicly made precisely this demand. Here, for example, the possibility of a "second wave" would have had to be discussed and requirements formulated for those responsible or the legislator as to how voluntariness can be permanently ensured.

Neither the fact that not all citizens have an app-enabled smartphone nor the fact that people could install an app on an old or second smartphone in order to prove the conditions of voluntary use are suitable arguments. Current discussions [9] about the CWA as an admission requirement or other benefits associated with it demonstrate the everyday relevance of this risk. Crucially, however, the requirements of consent must be present in each individual user, and so must voluntariness. A DPIA must address the question of how to assess a situation in which a majority or at least a large number of users can no longer be assumed to use the data voluntarily due to feelings of solidarity or employer coercion. At the very least, a legal regulation would have been a measure to be addressed. On this subject of the penetration of the population with CWA-capable devices and their voluntary use, two opposing statements can be found. Thus, with regard to voluntariness, it is stated: "In this respect, too, the voluntariness of the use of CWA could turn into a de facto compulsion through social pressure. However, it should be borne in mind that a significant proportion of the population do not own a smartphone at all, or do not own a suitable smartphone, especially if they are particularly young, old or have a low purchasing power." (Section 10.2.3.3 [15]) In the section on suitability, there is again a contrary expectation: "It is assumed that a large proportion of the population owns and mostly carries a suitable smartphone and that BLE technology may in principle be suitable for carrying out sufficiently precise distance measurement for logging contacts in the context of risk identification." (Section 11.2.2 [15]) So what is the basic assumption? Either the assumption on voluntariness or the assumption on suitability must be dropped by the responsible party.

### 3.4 Open questions

Some of the aspects outlined here have been taken up by the official DPIA team and included into the current DPIA [15], but many other key data protection questions remain open that would have to be addressed and analyzed in a DPIA. How does the generation of TeleTANs for health offices and hotline work? What happens on the portal server? How does the registration token behave, does it exclude traceability via QR code? What settings can the CWA user make? Who is the actual person responsible, who is the commissioning party, who has which responsibilities based on instructions? Can the RKI simply become the controller without legal assignment? Why is personal processing reduced to "evaluation of personal data" when both have their own (legal and technical) meaning?

## 4      Conclusion

Data protection cannot be implemented exclusively by technology and therefore cannot be evaluated by pure technology analysis of the IT components. The existing risks that arise from the activities of the responsible parties and their contracted service providers must be identified along the entire chain of processing steps and protective measures to reduce them must be proposed, discussed and evaluated. In order to identify a benchmark for the quality: A DPIA report should itself comply with the principles of Article 5 GDPR and the data protection goals. The legal dispute must specify the requirements to which the legal, technical and organisational measures for risk minimisation are subsequently aligned. Such a discussion was missing for the critical points of responsibility, purpose limitation of processing, the existence of the use of consent as well as the voluntariness of the use and the proof of voluntariness. The initially evaded risk discussion lead to the fact that an essential measure for risk reduction for the data subjects, namely the enactment of a law that binds the controller and other interested parties to the app, was not even discussed. It is not the purpose of a DPIA to justify the processing of personal data in the context of a technical solution, as was done in particular in the assessment of the proportionality of the processing in the initial DPIA. Rather, it is the task of the DPIA to identify the risks arising from the processing, the safeguards taken to that end and, in particular, to focus on unaddressed risks. The present official DPIA made a big step in this right direction [15].